\documentstyle[prd,aps, 
tighten, 
preprint, 
multicol]{revtex}

\begin{document}
\setcounter{page}{1}
\draft
\preprint{KIAS-P97016, hep-ph/9712406}

\title{ Strong CP and ${\bf \mu}$ Problems in Theories with 
        Gauge Mediated Supersymmetry Breaking }

\author{Eung Jin Chun}

\address{ Korea Institute for Advanced Study, \\
          207-43 Cheongryangri-dong, Dongdaemun-gu,
          Seoul 130-012, Korea}


\maketitle

\begin{abstract}
We provide a simple solution to the $\mu$ and strong CP problems in the 
context of gauge mediated supersymmetry breaking.   The generic appearance 
of R symmetry in dynamical supersymmetry breaking is used to implement 
Peccei-Quinn symmetry.  Acceptable $\mu$ and $B$ terms as well as the 
large symmetry breaking scale are induced in the presence of 
nonrenormalizable interactions.  Cosmological consequences of this scheme 
turn out to yield  constraints on the PQ symmetry breaking scale and 
the number of the messenger/heavy quarks.  Complexity in introducing 
non-R Peccei-Quinn symmetry is contrasted with the case of R symmetry.

\end{abstract}


Low energy supersymmetry breaking could be transmitted by gauge interactions 
from a hidden sector where supersymmetry is broken by nonperturbative
dynamics.  After the construction of viable models of this type \cite{minimal}, 
many works have been devoted to look for their experimental signatures  
\cite{many}.  These models also have distinctive features in cosmology 
\cite{chun}.
One of the theoretical difficulties in low energy supersymmetry breaking 
is how to generate the Higgs mass parameter, that is, the $\mu$ term.
In the context of the conventional supergravity models\cite{nilles}, 
the origin of the $\mu$ term can be attributed to the low energy
supersymmetry breaking scale $m_{3/2}$ which should be of the order of $\mu$.
On the other hand, the solution to the $\mu$ problem and the strong CP problem 
may have the same origin \cite{knc} in view of extending the Higgs sector 
\cite{dfsz}.  
The strong CP problem is solved elegantly by introducing Peccei-Quinn (PQ) 
symmetry \cite{pq} which is spontaneously broken at an intermediate 
scale $f_{PQ} \sim 10^{12}$ GeV.  This is also the scale of 
supersymmetry breaking at the hidden sector in gravity mediation scheme.

The $\mu$ problem in theories with gauge mediated supersymmetry breaking 
(GMSB) appears more severe.  Here the problem is to generate not only 
a right value of $\mu$ but also an acceptable $B$ term.
The existing proposals \cite{minimal,dgp,varia,gfm} typically suffer from 
problems of fine-tuning, or of complexity, etc.
Another difficulty in GMSB was the generic appearance of a spontaneously broken 
non-anomalous R symmetry in hidden sector models for dynamical supersymmetry 
breaking \cite{ns}.  Then the concern would be how to get rid of 
the dangerous R axion which is light and has a too low decay constant, $f_R$.  
Several solutions suggested in literature rely on lifting the mass of 
R axion \cite{bpr,asaka}.
First, R symmetry may be anomalous with respect to the hidden 
sector gauge group which will give mass of order 
$m_a \approx \Lambda_D^2/f_R$ where $\Lambda_D$ is the
dynamical scale of the hidden sector.
Second, R symmetry could be explicitly broken by the
constant which is introduced to tune the cosmological constant to zero,
in which case $m_a^2 \approx \Lambda_D^5/f_R^2 M_P$ \cite{bpr}.
Another way is to introduce a slight breaking of R symmetry
which is most desirable in view of solving the 
domain wall problem \cite{asaka}.
However, it would be more natural for this R axion to be 
the QCD axion \cite{bkn} keeping  the R symmetry exact 
and spontaneously broken 
by a field with a large vacuum expectation value (VEV)  $\sim f_{PQ}$.

In this paper, we propose to solve the $\mu$ problem in connection with
the strong CP problem in various schemes of GMSB.
For this purpose, we make use of nonrenormalizable interactions  
and  R symmetry generically appearing in GMSB models.
Nonrenormalizable interactions are necessary to generate the
spontaneous PQ symmetry breaking at the scale $f_{PQ}$. 
An important constraint on this scheme comes from the consideration of
domain wall problem keeping an exact R symmetry (up to anomaly).
Being PQ symmetry, R symmetry turns out to be very restrictive and 
constrains the theory, and thus more favorable than non-R symmetry.

\bigskip

In GMSB models, supersymmetry breaking in a hidden sector 
is mediated to the supersymmetric standard 
model (SSM) sector due to the presence of the following interaction
\begin{equation} \label{Wmess}
W_{mess} = kS f \bar{f} \,.
\end{equation}
Here $f,\bar{f}$ are vectorlike (messenger) quarks and leptons. 
The soft masses of ordinary squarks, sleptons and gauginos  are determined 
by the quantity $\Lambda_S = \langle F_S \rangle / \langle S \rangle$ 
which is induced by dynamical supersymmetry breaking in a hidden sector 
and should be $(10-100)$ TeV.
In order to generate the $\mu$ term, one needs to
extend the Higgs sector typically requiring more singlet fields.  
Our proposal here is to use a radiative mechanism which provides not only
the $\mu$ term but also a large PQ symmetry breaking scale \cite{msy,cck}.
For this one needs a singlet field $\phi$ which couples not only to 
the Higgs fields but also to vectorlike quark fields, $Q, \bar{Q}$, and has
nonrenormalizable interactions. 
The whole superpotential should respect the PQ (R) symmetry and 
takes the form,
\begin{equation} \label{WmuQ}
  W_\mu = {h_\mu \over m} H_1 H_2 {\phi^{m} \over M_P^{m-1}} + 
 {h_\phi\over n+3} {\phi^{n+3} \over M_P^n} + h_Q \phi Q \bar{Q} \,.
\end{equation}
The last term is reminiscent of heavy quark axion model \cite{hq} 
and was considered as a solution to the $\mu$ problem together with 
the trilinear term $\phi^3$ in Ref. \cite{dn}.  
The first term in Eq.~(\ref{WmuQ}) has been used also in Ref.~\cite{knc,dnns}.
The second term fixes the R symmetry of $\phi$.
It is the last term that generates a negative mass-squared  mass
($-m^2_\phi$) to the field $\phi$ through loop effect 
once $\langle \phi \rangle < \langle S\rangle $.  
Since the heavy squarks get the soft masses in the same way as 
ordinary squarks, the radiative mass of $\phi$ generated 
from tadpole diagram is given by
\begin{equation} \label{mphiII}
 m^2_\phi \simeq {3 N_Q h^2_Q \over 4 \pi^2} 
 \left(\ln{h_Q \langle \phi \rangle \over k\langle S \rangle }
         \right) m^2_{\tilde{q}} \,,
\end{equation}
where $N_Q$ is the number of the extra heavy quarks $(Q+\bar{Q})$.
Due to the above negative mass-squared, 
the superpotential (\ref{WmuQ}) produces a nonzero $\langle \phi \rangle$:
\begin{equation} \label{vevphi}
 \langle \phi \rangle = 
 \left( m_\phi M_P^n \over h_\phi \sqrt{n+2} \right)^{1/n+1}
\end{equation}
as well as the $\mu$ and $B$ terms:
\begin{eqnarray} \label{muB}
 \mu &=& {h_\mu \over m} 
 \left(m_\phi^{m} M_P^{n-m+1} \over h_\phi \sqrt{n+2} \right)^{1/n+1}
       \nonumber \\
 B &=& m_\phi \left(m \over \sqrt{n+2}\right)  \,.
\end{eqnarray}
%
In order to have $\mu \sim B$ without fine-tuning $h_\mu$, one needs $m=n+1$
and therefore,
\begin{eqnarray} \label{muBII}
 B &\simeq& m_{\phi}   \nonumber \\ 
 \mu &=& m_\phi \left( h_\mu \over m \right) 
                 \left(1 \over h_\phi\sqrt{m+1} \right)^{1/m} 
\end{eqnarray}
which are all of the order of the Higgs mass $m_{H_2}$.  
In this case, the value of $tan\beta$ is naturally of order 1.  
But the fine-tuning of $1\sim 6$ \% 
between the values of $m_{H_2}$ and $\mu$ cannot be avoided 
to get the right value of $Z$ boson mass \cite{ag}.  

For the radiative generation of a large VEV $\langle \phi \rangle$, 
the first term in Eq.~(\ref{WmuQ}) is necessary for the cosmological reason 
as well.  Since the PQ field $\phi$ is an almost flat direction
and gets a negative mass-squared, it drives 
so-called thermal inflation \cite{lyst}.  When the temperature of 
the universe cools down below the value $T \sim m_\phi$,
the VEV of the PQ field develops as given in Eq.~(\ref{vevphi}).
Then the vacuum energy of order $m_\phi^2 \langle \phi \rangle^2$ 
stored in the PQ field turns into oscillating matter energy 
dominating the universe.
Here, the concern is whether the subsequent decay products of the
(scalar part of) $\phi$ are sufficiently thermalized.
This can be achieved if $\phi$ has a sizable coupling with the ordinary
matter.  The $\mu$ term as in Eq.~(\ref{WmuQ}) indeed provides
effective thermalization of $\phi$ through its decay into 
two light Higgses or stops, dominating the decay into 
(non-thermalizable) two axions \cite{chlu}.
Having the $\phi$ decay rate, $\Gamma_\phi \approx
{1\over 16\pi} {m_\phi^3 \over \langle\phi\rangle^2}$, 
the limit on the reheat temperature $T_R > 6$ MeV after thermalization
translates into the limit $\langle\phi\rangle \lesssim 10^{15}$ GeV and
thus $m\leq 5$ [see Eq.~(\ref{phiF})].  
Let us recall also that  the axion whose population
is diluted by thermal inflation can be dark matter for $m=3,4$ \cite{cck}.
Hereafter, we consider only the case with $m\leq 5$.

Recall that this mechanism can work only if the messenger scale 
$\sim \langle S \rangle$ is higher than the PQ scale 
$\sim \langle \phi \rangle$, therefore can be used 
in the direct mediation models \cite{direct}.  
In this class of models, the field $S$ belongs directly to a hidden sector
and its VEV can be much larger than $\Lambda_S$.  
Typical models of direct mediation possess a rather large 
$\langle S \rangle \lesssim 10^{16}$ GeV.
To be precise, then, the PQ scale is determined by the large VEV
between  $\langle S \rangle$ and $\langle \phi \rangle$ 
if $S$ is also charged under the R symmetry.
In the below, we will show  an example of direct mediation models 
in which $S$ is neutral under the R symmetry.

\medskip

In the scheme under consideration, GMSB models get an important
restriction coming from cosmological consideration.
If the axion potential due to QCD instanton has degenerate vacua, 
domain walls can be produced \cite{sik} to cause overclosure of
the universe.
Hence, one has to inspect the vacuum structure of the axion 
potential for given models \cite{gw}.
The vacuum degeneracy is determined by the domain wall number $N_{DW}$
which is nothing but the QCD anomaly of PQ symmetry \cite{kim}.
If $N_{DW}=\pm 1$, there is no vacuum degeneracy and thus no domain wall 
problem.  
Let the R charges of $H_1$ and $H_2$ be $h_1$ and $h_2$, respectively,
under the usual R charge normalization $R(W)=2$.
The fermionic components carry the R charges one less than the bosonic 
components.
From the superpotentials (\ref{Wmess}) and  (\ref{WmuQ}), 
the R charges of the fermion bilinear operators are found to be 
\begin{equation}
 \begin{array}{cccc} 
  q\bar{u}& q\bar{d} & Q\bar{Q} &  f\bar{f}  \\ 
  -h_2 &  -h_1 &  -{2\over m+2} & -r_S 
 \end{array}
\end{equation}
where $h_1+h_2=4/(m+2)$.   Recall that the `Higgs fields', $H_2, H_1, 
\phi, S$, carry the R charges whose signs are opposite to those of the
corresponding fermion bilinear. The R charge $r_S$ of $S$ is determined
by the hidden sector in a given direct mediation model.
Now the domain wall number $N_{DW}$ is given by  
\begin{equation} \label{ndw}
 N_{DW} ={\cal N} [6 - {4 \over m+2} N_g -  {2 \over m+2} N_Q - r_S N_f] 
\end{equation}
where $N_Q$ is the number of the heavy quarks.  
The number ${\cal N}$ is a proper normalization factor 
under the condition that all fields carry integral charges 
(except the $Z_2$ ambiguity for fermion fields) and 
the common divisor is one. 
The normalization factor is determined after including the hidden sector
R charges (and thus $r_S$), and the condition $N_{DW}=\pm 1$ 
gives a fairly model-dependent restriction.
As an example, let us take the model discussed by Arkani-Hamed et.~al.\
in Ref.~\cite{direct}.  This model has $SU(7)\times SU(6)$ hidden sector
gauge group with particle contents
\begin{equation}
 \begin{tabular}{c|ccccccc} 
    &  $Q$ &  $L_i$ & $L_6$ & $R^i$  & $R^6$  & $R^7$ & $\phi^i$ \\ \hline 
 $SU(7)\times SU(6)$ &
     (7,6) & ($\bar{7}$,1) & ($\bar{7}$,1) & (1,$\bar{6}$) & (1,$\bar{6}$) & 
       (1,$\bar{6}$) & (1,1)  \\
 $SU(5)$ & 1 & $\bar{5}$ & 1 & 5 & 1 & 1 & 5 \\
 $U(1)_R$ & $-2$ & 4 & $-10$ & 0 & 14 & 2 & $-14$ \\
 \end{tabular}
\end{equation}
Here $Q,L_i$ contain 7 pairs of $(5+\bar{5})$ under the standard gauge group
$SU(5)$ and thus $N_f=7$.  The field $R^i$ play role of $S$,
and $\langle S \rangle \simeq 8\times 10^{14}$ GeV, 
$\langle F_S \rangle \simeq (3\times10^9 {\rm GeV})^2$.
Since $r_S=0$ in this example, the normalization factor can be ${\cal N}
=(m+2)/2$, and thus the domain wall number is $N_{DW}=3m-N_Q$.
To avoid the domain wall problem, therefore, the number of the extra 
heavy quarks has to be $N_Q=3m\pm1$. 
The condition that $\langle \phi \rangle  < \langle S \rangle$
gives a restriction $m\leq4$ [see Eq.~(\ref{phiF})].
One can find that perturbative unification is achieved
even if there are many extra quarks, $7+N_Q$.
Note also that the PQ symmetry breaking scale is given by
$f_{PQ} =\langle \phi \rangle$  as $r_S=0$.

\medskip 

Let us now briefly comment on the possibility of having non-R PQ symmetry
by relying on the radiative mechanism.
For this, one needs at least two fields, and both of them have to couple to
the heavy quarks in order to obtain nonzero VEVs. Therefore, one would have
\begin{eqnarray} \label{WnonR}
 W_{ren} &=& h_Q \phi_1 Q \bar{Q} + h_{Q'} \phi_2 Q' \bar{Q}' \\
 W_{\mu} &=& h_\mu H_1 H_2 {\phi_1^s \phi_2^t \over M_P^{s+t-1}} +
  h_\phi {\phi_1^p \phi_2^q \over M_P^{p+q-3}}\ . \nonumber
\end{eqnarray}
Here $p+q=s+t+2=m+2$ ($m=2-5$) is required as in the case of R symmetry.
Without loss of generality, we assign the PQ charges $-q$ and $p$ to 
$\phi_1$ and $\phi_2$, respectively.  In this case, only the SSM fields 
are charged under PQ symmetry.  The corresponding domain wall number is 
\begin{equation} \label{ndwnonR}
 N_{DW} = {1\over N}[(-sq+tp)N_g + q N_Q -p N_{Q'}]
\end{equation}
where the normalization factor is $N=1,2$ depending on the cases.
It becomes now clear that one could find many combinations of $(p,q,s,t)$
satisfying  $N_{DW}=\pm 1$, and also consistent with perturbative unification.
Compared to the R symmetry case, one may need only a smaller number 
of heavy quarks here.  
For suitable choices of $(p,q,s,t)$, the smallest possible 
number of the heavy quarks is $N_Q+N_{Q'}=3$.
At any rate, implementing non-R PQ symmetry needs more complication, and 
constrains less the models than R symmetry.

\bigskip

The above mechanism cannot work in the so-called minimal models \cite{minimal}
in which $\Lambda_S$ is generated by the presence of the couplings
\begin{equation} \label{Wmin}
 W'_{mess} = kSf\bar{f} + k'S^3 + k'' S\phi_+\phi_-
\end{equation}
where the singlet $S$ and its F term get VEVs through supersymmetry 
breaking effect delivered by the fields 
$\phi_{\pm}$ charged under a messenger gauge group $U(1)_m$.
Since $\Lambda_S \simeq (10-100)$ TeV, it is expected that 
$\Lambda_S \sim \langle S \rangle \sim \langle F_S \rangle$.
Furthermore, the supersymmetry breaking scale $\sqrt{F}$ 
in a hidden sector would be around $(10^5-10^7)$ GeV without fine-tuning.
However, one can still have a large VEV for a certain field in the hidden
sector again if nonrenormalizable terms are present, which could be 
related to the generation of  the $\mu$ term. 
To see this, let us suppose that a $m$-dimensional
operator ${\cal O}_{(m)}$ from the hidden sector couples to
the Higgs bilinear operator:
\begin{equation} \label{Wmu}
 W_{\mu} = \lambda_\mu {{\cal O}_{(m)} \over M_P^{m-1}} H_1 H_2 \,.
\end{equation}
Representing ${\cal O}_{(m)}$ in terms of a single field $\phi$, that is
${\cal O}_{(m)} \sim\phi^m$, the orders of magnitudes of $\mu$ and $B$ 
are given by 
\begin{equation}
 \mu \sim {\langle \phi^m \rangle \over M_P^{m-1} }\,,  \qquad 
 B \sim {\langle F_\phi \rangle \over  \langle\phi\rangle} \,.
\end{equation}
In order to have $\mu \sim B$, one needs $F_\phi \sim \phi^{m+1}$.  
Since $\langle F_\phi \rangle=F$, 
the tree-level superpotential of the hidden sector 
has to be of dimension $m+2$, that is, $W_{hidden} ={\cal O}_{(m+2)}
\sim \phi^{m+2}$.
From the above relations, we find the vacuum expectation values of 
a field $\phi$ and the supersymmetry breaking scale $\sqrt{F}$ in the
hidden sector: $\langle \phi \rangle \sim M_P(\mu/M_P)^{1/m}$,
$\sqrt{F} \sim M_P(\mu/M_P)^{m+1/2m}$. Taking $\mu= 100$ GeV, 
the approximate values of $\langle \phi \rangle$ and $\sqrt{F}$ in units
of GeV are
\begin{equation} \label{phiF}
 \begin{tabular}{c|cccc} 
  $m$ & 2 & 3 & 4 & 5 \\ \hline
  $\langle\phi\rangle$ & $2\times10^{10}$ & $8\times10^{12}$ &
        $2\times10^{14}$ & $1\times10^{15}$  \\
  $\sqrt{F}$ & $1\times10^{6}$ & $3\times10^7$ & $1\times10^8$ & 
              $4\times10^{8}$ \\ 
 \end{tabular}
\end{equation}
As mentioned, the hidden sector superpotential typically carries
a non-anomalous R symmetry, which is now broken at the scale $\langle 
\phi\rangle$ given above.  Due to the presence of the coupling (\ref{Wmu})
dictating the R charges of the ordinary fields,
this R symmetry becomes anomalous with respect to the ordinary QCD and
thus is a PQ symmetry, as a consequence of which  one gets
$f_{PQ}=\langle \phi \rangle$.  
It is interesting to observe that the restriction $m\leq3$ 
has to be imposed in view of the naturalness condition
 $\sqrt{F} \lesssim 10^7$ GeV  as well as the loose cosmological bound
$f_{PQ} \lesssim 10^{13}$ GeV.
If some mechanism to dilute the population of the coherent axion exists
and a slight fine-tuning is made, $m=4,5$ would be also acceptable.

\medskip

It is straitforward to read off the R (PQ) charges of the fields
from Eqs.~(\ref{Wmin}) and (\ref{Wmu}).  
Letting  the R charges of $H_1$ and $H_2$ be $h_1$ and $h_2$, respectively,
${\cal O}_{(m)}$ has the R charge $r_m=2-h_1-h_2$, 
and $S$ has the R charge $2/3$.
Then the fermion bilinear operators carry the R charges,
\begin{equation}
  \begin{array}{ccc} 
  q\bar{u} & q\bar{d} & f\bar{f} \\
  -h_2 & -h_1 &  -{2\over3} 
 \end{array}
\end{equation}
where $q, \bar{u}, \bar{d}$ denote the standard model quarks.
In order to know the proper normalization of the R charges again, 
one has to take into account the whole structure of the theory 
which should respect  the R selection rule.
That is, the R charges of all the fields appearing not only in the SSM or 
messenger sector but also in the hidden sector have to be known.  
At the moment, let us forget about the hidden sector to simplify 
the calculation.  
Under the proper normalization of the R charges, 
the domain wall number $N_{DW}$  becomes 
\begin{equation} \label{NDW1}
N_{DW}= {\cal N} [6-(h_1+h_2)N_g -{2\over3}N_f]
\end{equation}
where the normalization factor ${\cal N}$ can be $3$ or $3/2$
depending on the specific charge assignment, $N_g=3$ is the
number of ordinary quark generations, and 
$N_f$ is the number of messenger quarks.
Perturbative unification requires $N_f \leq 5$.  
The requirement $N_{DW}=\pm1$ can be satisfied in the following cases:
\begin{equation} \label{rmNf}
 (r_{(m)}, N_f) = ({1\over3}, 1 \mbox{ or } 2),\; 
   ({2\over3}, 2 \mbox{ or } 4), \; (1, 4 \mbox{ or } 5),\; 
   ({4\over3},5) \,. 
\end{equation}
For $r_{(m)}=1$, only $m=2$ is allowed since 
$W_{hidden}={\cal O}_{(m)}^2$ has to be of dimension $m+2$.

To illustrate the properties discussed above, let us take the well-known 
3-2 model \cite{3-2} which provides also the simplest 
realization of our scheme.
This model has the particle content with gauge group $SU(3)\times SU(2)$:
$Q=(3,2)_{1/3}$, $\bar{U}=(\bar{3},1)_{-4/3}$, 
$\bar{D}=(\bar{3},1)_{2/3}$, $L=(1,2)_{-1}$, $\bar{E}=(1,1)_{2}$.  
Here the subscripts denote the charges under 
the messenger group $U(1)_m$.  Allowing the hidden sector superpotential 
of dimension 6: $W_{hidden}=\lambda (LQD)^2$, 
one needs an operator ${\cal O}_{(m)}$
of dimension $m=4$ to produce right $\mu$ and $B$ terms as shown before.
Among dimension 4 operators, only $LQ\bar{U}\bar{E}$ 
can be used since the other operators $Q\bar{U}Q\bar{D}$, $QQQL$
cannot be compatible with hidden sector anomaly-free conditions.
The R charge assignment for the hidden sector have a freedom that
the  R charge of, e.g., $Q$ is not specified.  
Therefore, the normalization taken in Eq.~(\ref{NDW1}) can be kept 
for an appropriate choice of the R charge of $Q$, and any R charge 
except 1 in Eq.~(\ref{rmNf}) can be assigned to $LQ\bar{U}\bar{E}$.  
Usually, there are enough degrees of freedom 
in dynamical supersymmetry breaking models  to choose a 
R charge assignment which is free from the hidden sector anomaly.
If a hidden sector model has, however, no degrees of freedom in 
fixing the R charges, then certain model may not able to satisfy
$N_{DW}=\pm1$ and thus suffers from the domain wall problem 
in promoting  R symmetry to PQ symmetry.

\bigskip

In conclusion,  we showed that, while providing a solution to the $\mu$ and
$B$ problems in gauge mediated supersymmetry breaking models,
the strong CP problem can also be resolved by promoting R symmetry 
to PQ symmetry.  This can be done if a large symmetry breaking 
scale is generated in tree level or radiatively in the presence of 
nonrenormalizable interactions.  
It would be natural for R symmetry to be PQ symmetry as 
dynamical supersymmetry breaking models usually possess non-anomalous
R symmetry.
The cosmological consequences of this scheme (cosmological bound on the
PQ symmetry breaking scale, a late-time entropy production and 
the problem of domain wall formation) tightly restrict the models, 
which follows basically from the property of R symmetry.
Therefore, our scheme can be only realized in a certain gauge mediation model
or hidden sector model.

In the minimal type of models, 
the right values of $\mu$ and $B$ terms can be obtained  
if the $\mu$ term is generated by an operator of dimension $m$
while the hidden sector superpotential has dimension $m+2$.
Avoiding fine-tuning in these models, one finds restriction:
$m\leq 3$, which is also consistent with the cosmological bound on the
PQ scale $f_{PQ} \lesssim 10^{13}$ GeV.  A larger $m$ may be allowed 
if a late-time entropy production dilutes the (R) axion population.

The radiative generation of the PQ scale and $\mu$ term can be employed
in direct mediation models once the mediation scale is higher than the 
PQ scale.  For this, the introduction of extra heavy quarks 
which couple to a PQ singlet is needed. The $\mu$ term and a large PQ scale
can come from the nonrenormalizable coupling of the singlet field.
This mechanism provides also a late-time (thermal) inflation and thus
allows for a larger value of the PQ scale, $f_{PQ} \lesssim 10^{15}$ GeV.  
This puts a bound on the  dimensionality of the $\mu$ term: $m\leq 5$.
The sizes of $\mu$ and $B$ terms are determined to be of the  order of 
the radiative mass of the singlet which is of the order of $m_{H_2}$, 
and thus $tan\beta$ of order 1 is expected.  

In order to avoid the domain wall problem, the numbers of the messenger
fermions or heavy quarks have to be arranged consistently with
the R charge assignment to the $\mu$ term.  
This condition is model-dependent and two examples are worked out
for illustration.
In the minimal model with the $SU(3)\times SU(2)$ gauge group in the 
hidden sector, the composite $\mu$ operator has dimension 4 and the 
number of messenger fermions cannot be 3 which is also the case 
in generic minimal models. 
In the direct mediation model by Arkani-Hamed et.al., the radiatively 
generated  $\mu$ term can have dimension $m\leq 4$, 
and the number of heavy quarks has to be $N_Q=3m\pm1$.
We recall that, for the $\mu$ term of dimension 3, 
the PQ symmetry breaking scale is $f_{PQ} \approx 10^{13}$ GeV 
for which the corresponding axion 
can still be a candidate of cold dark matter.

In order to implement non-R PQ symmetry through the radiative mechanism, 
one needs to introduce at least two singlets with different PQ charges, 
and there are many possible PQ charge assignments 
which are free from the domain wall problem.
Thus non-R symmetry being PQ symmetry needs more complication
and constrains less the models than R symmetry.

\bigskip

{\bf Acknowledgments}:  
The author is supported by Non Directed Research Fund of 
Korea Research Foundation, 1996.  
He thanks ICTP for its hospitality during his stay for 
the Extended Workshop on Highlights in Astroparticle Physics
where this work has been developed.


\begin{thebibliography}{99}
\bibitem{minimal}
 M. Dine and A. Nelson, Phys. Rev. D {\bf 48}, 1277 (1993); 
 M. Dine, A. Nelson and Y. Shirman, Phys. Rev. D {\bf 51}, 1362 (1995);
 M. Dine, A. Nelson, Y. Nir and Y. Shirman, Phys. Rev. D {\bf 53}, 2658 (1996).
\bibitem{many} 
  S. Dimopoulos, M. Dine, S. Raby and S. Thomas, Phys.\ Rev.\ Lett.\ {\bf 76}, 
  3494 (1996); D. R. Stump, M. Wiest, C. P. Yuan, Phys.\ Rev.\ D {\bf 54}, 
  1936 (1996); S. Ambrosanio, G. L. Kane, G. D. Kribs, S. P. Martin and 
  S. Mrenna, Phys.\ Rev.\ Lett.\ {\bf 76}, 3498 (1996); Phys.\ Rev.\ D 
  {\bf 54}, 5395 (1996); 
  S. Dimopoulos, S. Thosmas and J. D. Wells, Phys.\ Rev.\ D {\bf 54}, 
  3283 (1996);  
  K. S. Babu, C. Kolda and F. Wilczek, Phys.\ Rev.\ Lett.\ {\bf 77}, 
  3070 (1996);  
  K. Kiers, J. N. Ng and  G. Wu, Phys.\ Lett.\ B {\bf 381}, 177 (1996); 
  J. L. Lopez and D. V. Nanopoulos, Phys.\ Rev.\ D {\bf 55}, 4450 (1997); 
  H. Baer, M. Brhlik, C.-h. Chen and X. Tata, Phys.\ Rev.\ D {\bf 55}, 
  4463 (1997); 
  A. Ghosal, A. Kundu and B. Mukhopadhyaya, Phys.\ Rev.\ D {\bf 56}, 
  504 (1997); 
  D. A. Dicus, B. Dutta and S. Nandi, Phys.\ Rev.\ Lett.\ {\bf 78}, 
  3055 (1997); 
  A. Ambrosanio, G. D. Kribs and S. P. Martin, Phys.\ Rev.\ D {\bf 56}, 
  1761 (1997).
\bibitem{chun}
  E. J. Chun, H. B. Kim and J. E. Kim, Phys.\ Rev.\ Lett.\ {\bf 72}, 
  1956 (1994); 
  A. de Gouvea, T. Moroi and H. Murayama, Phys.\ Rev.\ D {\bf 56}, 1281 (1997); 
  J. Hashiba, M. Kawasaki and T. Yanagida, {\it preprint} hep-ph/9708226 
   (to appear in Phys.\ Rev.\ Lett.).
\bibitem{nilles} 
  For a review, see H. P. Nilles, Phys.\ Rept.\ {\bf 110}, 1 (1984).
\bibitem{knc} 
  J. E. Kim and H. P. Nilles, Phys.\ Lett.\ B {\bf 138}, 150 (1984);
  E. J. Chun, J. E. Kim and H. P. Nilles, Nucl.\ Phys.\ B {\bf 370}, 105 (1992).
\bibitem{dfsz} 
  A. R. Zhitnitskii, Sov.\ J. Nucl.\ Phys.\ {\bf 31}, 260 (1980);
  M. Dine, W. Fischler and M. Srednicki, Phys.\ Lett.\ B {\bf 104}, 199 (1981).
\bibitem{pq} 
  R. D. Peccei and H. R. Quinn, Phys.\ Rev.\ Lett.\ {\bf 38}, 1440 (1977); 
   Phys.\ Rev.\ D {\bf 16}, 1791 (1977).
\bibitem{dgp} 
  G. Dvali, G. F. Giudice and A. Pomarol, Nucl.\ Phys.\ B {\bf 478}, 31 (1996).
\bibitem{varia} 
  M. Dine, Y. Nir and Y. Shirman, Phys.\ Rev.\ D {\bf 55}, 1501 (1997).
\bibitem{gfm} 
  A. de Gouvea, A. Friedland and H. Murayama, {\it preprint} hep-ph/9711264.
\bibitem{ns} 
  A. E. Nelson and  N. Seiberg, Nucl.\ Phys.\ B {\bf 416}, 46 (1994).
\bibitem{bpr} 
  J. A. Bagger, E. Poppitz and L. Randall, Nucl.\ Phys.\ B {\bf 426}, 3 (1994).
\bibitem{asaka}
  T. Asaka,  J. Hashiba, M. Kawasaki and T. Yanagida, hep-ph/9711501;
  E. D. Stewart, M. Kawasaki and T. Yanagida, 
       Phys.\ Rev.\ D {\bf 54}, 6032 (1996).
\bibitem{bkn}
  T. Banks, Kaplan, A. Nelson, Phys.\ Rev.\ D {\bf 49}, 779 (1994).
\bibitem{msy} 
  H. Murayama, H. Suzuki and T. Yanagida, Phys.\ Lett.\ B {\bf 291}, 418 (1992).
\bibitem{cck} 
  K. Choi, E. J. Chun and J. E. Kim, Phys.\ Lett.\ B {\bf 403}, 209 (1997).
\bibitem{hq} 
  J. E. Kim, Phys.\ Rev.\ Lett.\ {\bf 43}, 103(1979); M. A. Shifman, 
  V. I. Vainstein and V. I. Zakharov, Nucl.\ Phys.\ B {\bf 166}, 493 (1980) .
\bibitem{dn} 
  M. Dine and A. Nelson in Ref.~\cite{minimal}.
\bibitem{dnns} 
 M. Dine, A. Nelson, Y. Nir and Y. Shirman, in Ref.~\cite{minimal};
 M. Leuerer, Y. Nir and N. Seiberg, 
   Nucl.\ Phys.\ B {\bf 420}, 468 (1994).
\bibitem{ag} 
  K. Agashe, M. Graesser, {\it preprint}  hep-ph/9704206.
\bibitem{lyst} 
  D. H. Lyth and E. D. Stewart,  Phys.\ Rev.\ Lett.\ {\bf 75}, 201 (1995);
  Phys.\ Rev.\ D {\bf 53}, 1784 (1996).
\bibitem{chlu}  
  E. J. Chun and A. Lukas, Phys.\ Lett.\ B {\bf 357}, 43 (1995).
\bibitem{laza} 
  G. Lazarides, R. K. Schaefer, D. Seckel and Q.  Shafi, 
  Nucl.\ Phys.\ B {\bf 346}, 
  193 (1990);  G. Lazarides, C Panagiotakopoulos and Q. Shafi, Phys.\ Lett.\ 
  B {\bf 192}, 323 (1987).
\bibitem{direct} 
  E. Poppitz and S. P. Trivedi, Phys.\ Rev.\ D {\bf 55}, 5508 (1997);
  N. Arkani-Hamed, J. March-Russell and H. Murayama, {\it preprint} 
  hep-ph/9701286;
  N. Haba, N. Maru and T Matsuoka, Phys.\ Rev.\ D {\bf 56}, 4207 (1997);
  H. Murayama, Phys.\ Rev.\ Lett.\ {\bf 79}, 18 (1997).
\bibitem{sik} 
  P. Sikivie, Phys.\ Rev.\ Lett.\ {\bf 48}, 1156 (1982).
\bibitem{gw} 
  H. Georgi and M. B. Wise, Phys.\ Lett.\ B {\bf 116}, 123 (1982).
\bibitem{kim} 
  For  reviews see, J. E. Kim, Phys.\ Rep.\ {\bf 150} (1987) 1;
  E. W. Kolb and M. S. Turner, {\it The Early Universe} (Addison-Wesley, 1990).
\bibitem{3-2} 
  I. Affleck, M. Dine and N. Seiberg, Nucl.\ Phys..\ B {\bf 256}, 557 (1985);
  M. Dine, A. Nelson and Y. Shirman, in Ref.~\cite{minimal}.

\end{thebibliography}
\end{document}